\documentclass[fleqn,12pt,twoside]{article}
\usepackage{espcrc1}

\usepackage{graphicx}
\usepackage[figuresright]{rotating}

\newcommand{\AmS}{{\protect\the\textfont2
  A\kern-.1667em\lower.5ex\hbox{M}\kern-.125emS}}

\title{Low-lying  resonances of   $^9_\Lambda$Be: Faddeev calculation with
Pade'-approximants}
\author{I. Filikhin\address[nccu]{Department of Physics, North Carolina Central
University, 1801 Fayetteville Street, Durham, NC 27707, USA},
 V. M. Suslov\addressmark\thanks{Department of Mathematical and Computational Physics,
    Sankt-Petersburg State University, 198504 Ul'yanovskaya Str.1, Petrodvorets,
 Russia}
 and
 B. Vlahovic\addressmark\thanks {This work was supported by the
Department of Defense and NASA through Grants, W911NF-05-1-0502 and
NAG3-804, respectively.}}

\begin{document}

% typeset front matter
\maketitle

\begin{abstract}
Configuration space Faddeev equations are applied to describe the
$^9_{\Lambda}$Be hypernucleus in the $\alpha\alpha\Lambda$ cluster
model. For calculation of resonance state energies a variant of the
method of analytical continuation in coupling constant is used. To
realize this method, an auxiliary three-body potential is added to
the Hamiltonian of the equations. A proper choice of strength
parameter of the potential converts a resonance state into a bound
state and an energy trajectory is obtained by variations of this
parameter. To fulfill analytical continuation of this trajectory as
a function of the strength parameter onto the complex plane, the
Pade' approximation is used. Spectrum of low-lying resonances is
calculated with two $\alpha\Lambda$ phenomenological potentials. We
predict existence of the 0$^+_2$ and 4$^+_1$ virtual states as well
as the 2$^+_2$ resonance state near by the $\alpha\alpha\Lambda$
threshold. The $^8$Be($L^+$)+$\Lambda$($s$-wave) configuration for
description of the $^9_{\Lambda}$Be ground band is discussed.
\end{abstract}

\section{Introduction}
Recently a new interest appeared to the method of analytical
continuation in coupling constant. The method was suggested
\cite{KKH} a long time ago to calculate resonance state parameters
of a three-body system. Though this method is based on approximate
continuation onto the complex plane using a number of negative
energies, enough accurate results were obtained in \cite{KK}. In
that paper the analytical continuation was applied combined with
complex scaling method. In paper \cite{FSV}, it was also shown that
this method allows one to obtain accurate calculations of low-lying
resonances.

In the present work we applied a variation of the method of
analytical continuation in coupling constant for calculations of
energy of resonances using the $\alpha+\alpha+\Lambda$ cluster model
\cite{HT06} for the $^9_{\Lambda}$Be hypernucleus. In this variant
of the method an additional non-physical three-body potential is
defined. To perform the analytical continuation, we use a number of
negative energy values obtained by changing the strength parameter
(coupling constant) of this additional potential. The unchanging of
the pair interactions and two-body thresholds is an important
attribute of such an approach. Calculations of bound state energies
are performed on the basis of the Faddeev equations in the
configuration space. To describe $\alpha\alpha$ nuclear interaction,
the Ali-Bodmer potential \cite{ABo66} having $s$, $d$ and $g$-wave
components is used. Spectra of low-lying resonances are calculated
with two $\alpha\Lambda$ phenomenological potentials. We predict the
existence of a 2$^+$ resonance state close to the
$\alpha\alpha\Lambda$ threshold. The classification of the
$^9_{\Lambda}$Be ground band levels which was given in
\cite{HT06,MBI,Bando} is discussed.

\section{Model and method}

Coupled set of the Faddeev equations describing the
$_{\Lambda}^{~9}$Be nucleus in the $\alpha\alpha\Lambda$ cluster
model is written in the form:
\begin{equation}
\begin{array}{l}
(H_0+V^s_{\alpha\Lambda}+V^c-E)W=-V^s_{\alpha\Lambda}(U-P_{12}W),\\
(H_0+V^s_{\alpha\alpha}+V^c-E)U=-V^s_{\alpha\alpha}(W-P_{12}W),
\end{array}
\label{F}
\end{equation}
\noindent where $H_0$ is the kinetic energy operator, $P_{12}$ is
the permutation operator for the  $\alpha$ particles (particles
1,2), $V^s_{\alpha\alpha}$ and $V^s_{\alpha\Lambda}$ are nuclear
potentials of $\alpha\alpha$ and $\alpha\Lambda$ interactions,
respectively. $V^c$ is the potential of the Coulomb interaction
between $\alpha$ particles, $U$ is the Faddeev component
corresponding to the rearrangement channel $(\alpha\alpha)-\Lambda$
and $W$ corresponds to the rearrangement channel
$(\alpha\Lambda)-\alpha$. The total wave function is expressed via
the components $U$ and $W$: $\Psi = U + (1 - P_{12})W$. The total
orbital angular momentum is given by $\vec
L={\vec\ell}_{\alpha\alpha}+{\vec\lambda}_{(\alpha\alpha)-\Lambda}=
{\vec\ell}_{\alpha\Lambda}+{\vec\lambda}_{(\alpha\Lambda)-\alpha}$,
where ${\ell}_{\alpha\alpha}$ (${\ell}_{\alpha\Lambda}$) is the
orbital angular momentum of the $\alpha$'s (pair of
${\alpha\Lambda}$) and ${\lambda}_{(\alpha\alpha)-\Lambda}$
(${\lambda}_{(\alpha\Lambda)-\alpha}$) is the orbital angular
momentum of a $\Lambda$ hyperon ($\alpha$ particle) relative to the
center of mass of the pair $\alpha$ ($\alpha\Lambda$) particles.
More detailed description of this formalism may be found in ref.
\cite{FGS04}. To describe interactions in the $\alpha \alpha
\Lambda$ system, local pairwise potentials were used. The nuclear
$\alpha\alpha$ interaction is given by version "a" of the
phenomenological Ali-Bodmer potential \cite{ABo66} having $s$, $d$
and $g$ - wave components, which was modified in work \cite{FJ}. For
$\alpha\Lambda$ interaction, phenomenological potentials having the
form of one (Gibson \cite{Gibson}) and two rank (Isle \cite{KAT85})
Gaussian were used. Each potential have the same components as
$s$-wave component in all partial waves of the $\alpha\Lambda$ pair.
To estimate the energies of low-lying resonance states, we applied a
method \cite{KK,FSV} which is based on an analytical continuation
with strength parameter for additional three-body potential
(coupling constant). This continuation in the unbound state region
is carried out using the Pad\'{e} approximants. The three-body
potential having the form: $V_3(\rho) = -\delta\exp(-b\rho^2)$ is
added in the left hand sides of Eqs. (\ref{F}). Parameter of this
potential $b$=0.1$fm^{-1}$, $\delta\geq$0 is variational parameter,
and $\rho^{2}=x_{\alpha}^{2}+y_{\alpha}^{2}$ where
$x_{\alpha},y_{\alpha}$ are the scaled Jacobi coordinates
($\alpha=1,2$) \cite{FSV}. For each resonance there exists the
region $|\delta |\geq |\delta_0 |$ where a resonance is by a bound
state. The three-body bound state calculations were performed using
(2$N$) values of $\delta$. The continuation onto complex plane is
carried out by means of the Pad\'{e} approximant: $
   \sqrt{-E}=\frac{\sum_{i=1}^Np_i\xi^i}{1+\sum_{i=1}^Nq_i\xi^i}
$ where $\xi=\sqrt{\delta_0-\delta}$. The Pad\'{e} approximant for
$\delta=0$ gives the energy and width of resonance:
$E(\delta=0)=E_r+i\Gamma/2$. Note that accuracy of the Pad\'{e}
approximation for resonance energy and width decreases with
increasing of distance from scattering threshold.

\section{Calculations}
\begin{figure}[htb]
\begin{minipage}[t]{80mm}
\framebox[80mm] {\includegraphics[height=79mm,width=75mm]
{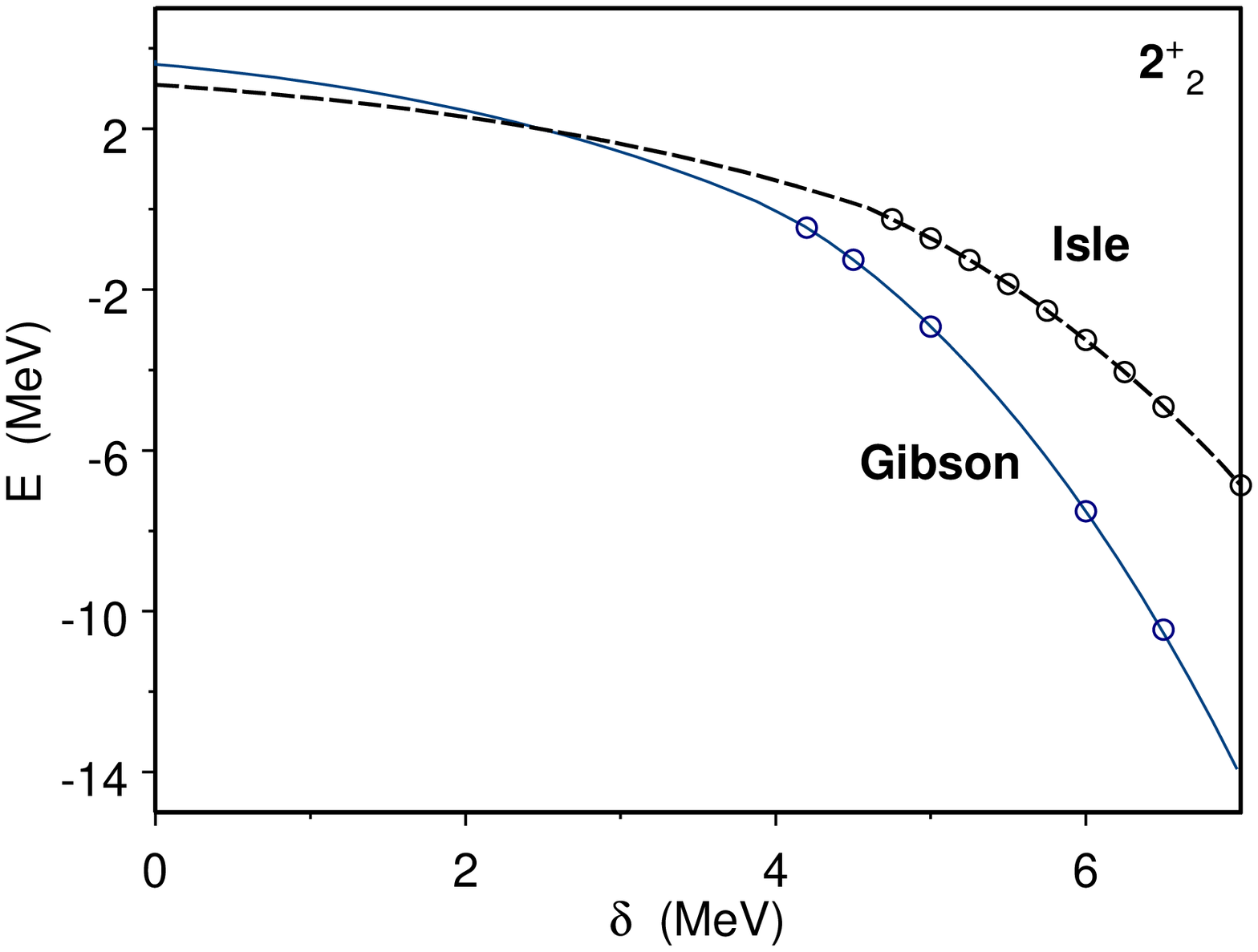} \rule[-1mm]{0mm}{73mm}} \caption{The Pade'
approximants for 2$_2^+$ resonance of the $\alpha\alpha\Lambda$
system. The solid (dashed) line corresponds to calculation with
Gibson(Isle) $\alpha\Lambda$ potential. Calculated bound state
energies are shown by circles. Energies are measured from the
$^5_\Lambda$He+$\alpha$ threshold.} \label{fig1}
\end{minipage}
\hspace{\fill}
\begin{minipage}[t]{80mm}
\framebox[80mm] {\includegraphics[height=79mm, width=78mm]{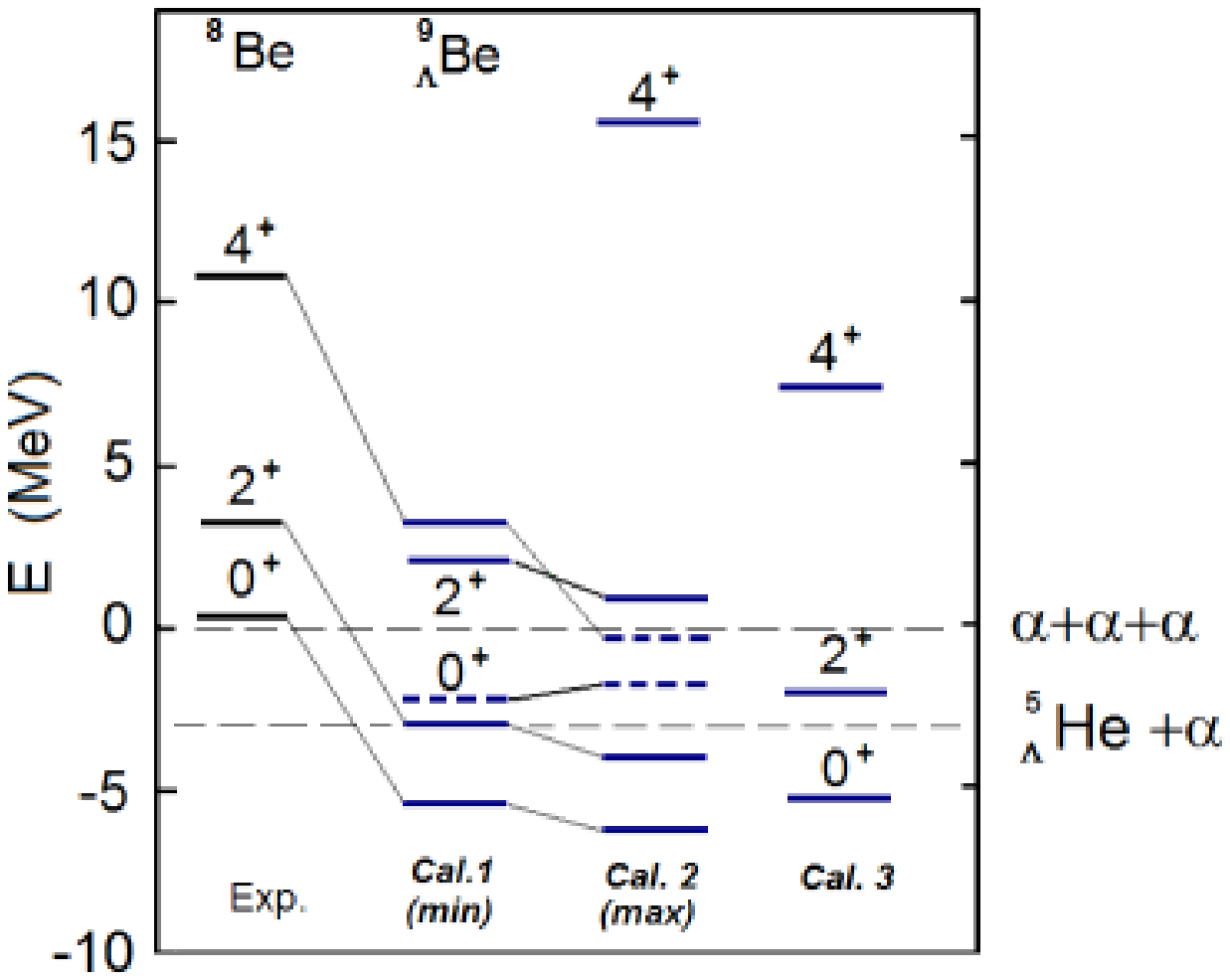}
\rule[-1mm]{0mm}{73mm}} \caption{Experimental data \cite{TUNL} for
energies of  $^8$Be and $^9_\Lambda$Be ground band levels calculated
with the Gibson $\alpha\Lambda$ potential. $ Cal.1(min) $
corresponds to the results of the calculations with "minimal"
configuration of orbital momenta. For $ Cal.2(max) $ the orbital
momenta are taken in "maximal" configuration. $Cal.3$ corresponds to
the results of \cite{Bando}. Virtual states are shown by dashed
lines.} \label{fig2}
\end{minipage}
\end{figure}

  The coupled configuration space
Faddeev equations (\ref{F}) for the $\alpha\alpha\Lambda$ system are
solved numerically using the finite-difference method and spline
appoximation \cite{FGS04,FGS}. In Table ~\ref{tab1} we present the
$2^+_2$ resonance parameters calculated with Pade' approximation of
various orders. Accuracy of our calculations is defined by the
decimal digit. Pade' approximant as a function of the "coupling
constant" $\delta$ is shown in Figure \ref{fig1} together with
values of bound state energies used for construction of the
approximants. The calculated energies of the resonances ($2^+_2$,
$4^+_2$), bound ($0^+_1$, $2^+_1$) and virtual ($0^+_1$, $4^+_1$)
states for $^9_{\Lambda}$Be ground band are shown in
Figure~\ref{fig2} for the case of the Gibson $\alpha\Lambda$
potential. For the Isle potential, the localization of low-lying
levels is similar to one showed in Figure~\ref{fig2}, though the
potential overbinds the ground state for 2~MeV, approximately
\cite{FGS04}. Results for Isle potential are shifted downwards
relative to corresponding Gibson potential results for the $0^+_1$,
$2^+_1$, $4^+_2$ states. In Figure~\ref{fig2}, we show also the
experimental data for $^8$Be ground band \cite{TUNL}. In ref.
\cite{HT06,MBI,Bando} ground band of $^9_{\Lambda}$Be was classified
as an analog of the $^8$Be ground band. According to this
classification $^9_{\Lambda}$Be is considered as the $^8$Be core
plus $s$-wave hyperon: $^8$Be(L$^+$)+$\Lambda$($s$-wave), when
$L=0,2,4$. As is seen from Figure~\ref{fig2}, the results of our
calculation do not agree with such simple scheme. In these
calculations we used two sets of subsystem momenta $l$ and $\lambda$
to evaluate contributions in energy due to higher partial waves. The
minimal configurations of orbital momenta
($l_{\alpha\alpha}$,$\lambda_{(\alpha\alpha)-\Lambda}$)
corresponding to the states $0^+$, $2^+$, $4^+$ are \{(0,0)\},
\{(0,2),(2,0)\}, \{(0,4),(4,0)\}, respectively. At the same time in
the $(\alpha\Lambda)-\alpha$ channel only the first configurations
with minimal values of momenta are used. Corresponding results in
Figure~\ref{fig2} are called "minimal" ones. Note that from Figure
\ref{fig2} one could conclude that other possible orbital momenta
configurations play an important role in formation of these states.
Higher partial waves  taken into account in orbital momenta
configurations shift the energy levels. In Figure~\ref{fig2} these
configurations are called "maximal" ones. They include  six first
minimal successive values of momenta for both
$(\alpha\alpha)-\Lambda$ and $(\alpha\Lambda)-\alpha$ channels. We
have found a new 2$^+_2$ resonance state that is near by the
$\alpha\alpha\Lambda$ threshold. In addition the 0$^+_2$ and 4$^+_1$
states are appeared as virtual states. From Figure~\ref{fig2} one
can see the $4^+_1$ state, which is a resonance state in minimal
configuration, becomes a virtual state in maximal configuration. The
$4^+$ resonance of $^9_{\Lambda}$Be is appeared as  the second $4^+$
state.

\begin{table}[htb]
\caption{Calculated 2$^+_2$ energy $E_r$ and width $\Gamma/2$
($E_r+i\Gamma/2$) (in MeV) in dependence on the order $N$ of the
Pade' approximant with $\delta_0$~=~-4.0~MeV (-4.62~MeV) for Gibson
(Isle) potential. The energy is measured from the
$^5_\Lambda$He+$\alpha$ threshold.} \label{tab1}
\newcommand{\m}{\hphantom{$-$}}
\newcommand{\cc}[1]{\multicolumn{1}{c}{#1}}
\renewcommand{\tabcolsep}{2pc} % enlarge column spacing
\renewcommand{\arraystretch}{1.2} % enlarge line spacing
\begin{tabular}{@{}lll}
\hline
$N$ & Gibson & Isle \\
\hline
 2     & 3.8(0)+i4.3(2)  & 3.2(0)+i3.2(6)\\
 3     & 3.8(2)+i4.3(2)  &  -- \\
 4     & 3.8(3)+i4.3(2)  & 3.0(9)+i3.2(6)\\
\hline
\end{tabular}\\[2pt]
\end{table}

\end{document}